\documentclass[useAMS,usenatbib]{mn2e}
\usepackage{graphics,epsfig,aas_macros}
\usepackage[normalem]{ulem}
\usepackage[dvipsnames]{color}
\usepackage[]{inputenc,amssymb}
\usepackage{amsmath}
\usepackage{color}

\usepackage[toc,page]{appendix}

\def \be{\begin{equation}}
\def \ee{\end{equation}}
\def \bea{\begin{eqnarray}}
\def \eea{\end{eqnarray}}
\def\etal{{et al.}}

\title[Suppression of galactic outflows]{
Suppression of galactic outflows by cosmological infall and circumgalactic medium}

\author[Singh \etal]
{Priyanka Singh$^{1}$\thanks{
priyankas@rri.res.in}, Sandeep Rana$^{2}$, Jasjeet S. Bagla$^2$, Biman B. Nath$^{1}$\\
    $^{1}$Raman Research Institute, Sadashiva Nagar, Bangalore, 560080, India\\
  $^{2}$Indian Institute of Science Education \& Research, Mohali, Mohali
}

\begin{document}

\date{Accepted, Received; in original form}
%\date{Accepted 2010 May 19;Received 2010 May 19; in original form 2010 February 13}

\maketitle

\label{firstpage}

\begin{abstract}
  We investigate the relative importance of two galactic outflow
  suppression mechanisms : a) Cosmological infall of the
  intergalactic gas onto the galaxy, and b) the existence of a hot
  circumgalactic medium (CGM). Considering only radial motion, the
  infall reduces the speed of outflowing gas and even halts the
  outflow, depending on the mass and redshift of the galaxy.  
  For star forming galaxies
  there exists an upper mass limit beyond which outflows are
  suppressed by the gravitational field of the galaxy.
  We find that infall can reduce this upper mass limit approximately
  by a factor of two (independent of the redshift).
  Massive galaxies ($\gtrsim 10^{12} M_{\odot}$) host large reservoir of
  hot, diffuse CGM around the central part of the galaxy.
  The CGM acts as a barrier between the infalling and outflowing gas and
  provides an additional source of outflow suppression.
  We find that at low redshifts ($z\lesssim3.5$), the CGM is more effective
  than the infall in suppressing the outflows.
  Together, these two processes give a mass range in which galaxies are
  unable to have effective outflows.
  We also discuss the impact of outflow suppression on the enrichment
  history of the galaxy and its environment.
\end{abstract}


\section{Introduction}

The two components of matter in galaxies-- dark matter and baryons--
have contrasting properties, a fact which makes the study of galactic
evolution a complex one.
Unlike dark matter, baryons undergo collisions, radiate, and condense
towards the central part of the galaxy in order to form
stars.
However, a significant fraction of these baryons may remain too hot to
condense and form stars, especially in large galaxies.
This gas likely stays in a hot, diffuse, gaseous form and envelopes
the central, optically visible part of the galaxy \citep{rees77,
  silk77}.
This gaseous component of the galaxy is referred to as the
circumgalactic medium (CGM).  
Recent observations have indicated the presence of the CGM in massive
galaxies \citep{anderson11, dai12, anderson13, bogdan13a, bogdan13b},
perhaps extending up to the virial radius of the galaxy.  

Galaxies also undergo feedback processes like supernovae (SNe) and
active galactic nuclei (AGN) which may give rise to galactic-scale
outflows \citep{croton06, dave11, vogel13}. 
Simultaneously, the galaxies also accrete matter from their
surrounding intergalactic medium (IGM) \citep{birn03, keres05,
  opp10}. 
Together, the infall and outflows regulate the evolution of the host
galaxies.
However, the interaction between these two opposing processes is not
yet well understood.   

The CGM can also act as a barrier for infalling gas as well as the
outflowing material if the CGM gas cooling timescale is comparable to
or larger than the halo destruction timescale \citep{singh15a} leading
to a hot, diffuse gaseous barrier between the central galaxy and IGM.
Recent results of observations \citep{mathes14} and simulations
  \citep{goerdt15, gabor15} can be explained by the  
existence of the hot CGM suppressing the outflows as well as infall.
The outflows are generally metal rich and remove a significant amount
of metal from galaxies.
Since the outflows can be decelerated and even stopped by the hot CGM
in massive galaxies, the recycling of metals in massive galaxies
becomes more important as compared to the low mass galaxies.  
This mass dependent recycling behaviour is known as the differential
wind recycling \citep{opp10} and it can alter the metal evolution
history of these galaxies and the surrounding IGM.
For low mass galaxies($M_h \lesssim 10^{12} M_{\odot}$), the gas
cooling timescale is small compared the halo destruction timescale.
As a result low mass galaxies cannot sustain the hot CGM gas.
This gas cools down, form clumps and does not interfere much with in
the infall/outflows.
Whereas, in case of massive galaxies, the CGM gas remains hot for long
enough timescale, leading to the existence of hot, gaseous barrier 
decelerating infalling \citep{dekel06} as well as outflowing material.  
Therefore, the presence of the hot CGM divides the galaxies into two
categories: 1). Massive galaxies ($M_h \gtrsim 10^{12} M_{\odot}$),
where the CGM is hot enough to affect the physical properties (infall,
outflows, metal enrichment etc.) of the galaxy and the wind recycling
becomes important, 2). low mass 
galaxies ($M_h \lesssim 10^{12} M_{\odot}$), where the CGM cools fast
enough and is essentially invisible to the outflowing and infalling
gas. 

In this work, we address the question of interaction between outflows,
infalling gas and CGM from a different perspective. 
Instead of focussing on the fate of infalling gas, we would like to
study the effect of infalling gas and CGM on the outflowing gas. 
Can outflowing gas escape to the IGM, or is its ultimate mixing with
the IGM suppressed?
How does this suppression, if at all, depend on the galactic mass and
redshift?
The possible suppression of outflows is more important for low mass
haloes where the hot CGM is essentially absent.
For a given halo mass, there exists a redshift where the suppression
of outflows by the presence of the hot CGM becomes more important than
its suppression by infall. 
The relative importance of the two wind suppression mechanisms depends
on the mass and redshift of the galactic halo. 
Therefore, it is important to take into account both the mechanisms to
understand the galaxy evolution and enrichment. 

However, the task is made a difficult one by the complications
inherent in the physics of outflowing gas, and also in the complicated
nature of infalling gas.
Firstly, the outflowing gas may not be spherically symmetric and may
have a complicated density, temperature and velocity structure, and
this structure itself may be a function of time.
Secondly, the infalling gas may also have an anisotropic density and
velocity structure.
One way to approach the problem is to use cosmological hydrodynamical
simulations, which, however, is unlikely to help in understanding the
physical processes involved, because of the complexity of the
processes.
The other approach is to set up idealised numerical experiments, in
which certain parameters are varied keeping the others constant, and
the processes are studied in detail.
However, even before such an exercise, it is useful to study idealised
theoretical scenarios with a mix of analytical and numerical tools.
This is what we attempt here.  
In this paper, for outflows, we use the analytical prescription by
\cite{sharma13}.  
For infall we use N-body simulations with TreePM code and $N=512^3$
particles \citep{bagla02, khandai09}.
Used together, they allow us to arrive at a few interesting
conclusions regarding the suppression of outflows by infalling matter
and CGM, which may have important implications in the cosmological
context. 

This paper is organized as follows: In section-\ref{sec-form} we
describe the formalism used to calculate the infall and outflow
velocities.
In section-\ref{sec-supp} we discuss the outflow suppression processes
and estimate the relative importance of these processes.
In section-\ref{sec-igm} we discuss the impact of the suppression on the 
IGM enrichment and present our main conclusions in
section-\ref{sec-conc}. 

\section{Formalism}
\label{sec-form}
\subsection{Outflow velocity}
\label{subs-vout}

The velocity of the outflowing gas mainly depends on the mass and
redshift of the galactic halo and the feedback recipe considered. 
\cite{sharma13} derived the terminal velocity of outflows driven by
multiple supernovae in a galaxy whose dark matter profile is described
by the Navarro-Frenk-White (NFW) profile.
They showed that the wind speed at large galacto-centric distance
depends on two velocity scales: (a) $v_\ast$, which depends on the
mass and energy deposition rate due to supernovae, and is given by
$v_\ast\approx (\dot{E} /2 \dot{M})^{1/2}$, and (b) $v_s$, which
depends on the dark matter profile, and is closely related to the
circular speed in a NFW profile. 
The terminal speed of winds (in the absence of momentum injection from
active galactic nuclei) was shown to be, 
\be
  v_{\rm{wind}}(r)=2\Bigl[ v^{2}_{*}-\frac{1}{2} \Bigl[\phi_{NFW}(r)-\phi_{NFW}(R)\Bigr]\Bigr] ^{1/2}
\ee
where R=200pc is assumed to be the sonic point, as well as the size of
the region in which mass and energy is being injected and
$\phi_{NFW}(r)=-2 v^{2}_{s} \frac{\ln(1+r/r_s)}{r/r_s}$ is the NFW
gravitational potential. 
The terminal wind velocity ($r\rightarrow\infty$) is given by:
\begin{equation}
 v_{\rm{term}}=2(v^{2}_{*}-v^{2}_{s})^{1/2}
 \label{terminal}
\end{equation}
The velocity scale inherent in the energy deposition is given by
$v_{*}=562\surd\alpha$ km/s, and it is due to effect of SNe, where
$\alpha$ represents energy injection efficiency. 
The other velocity scale, $v_{s}$ = $\surd\frac{GM_{h}}{Cr_{s}}$ is
due to the gravity of the halo, where, $M_h$ is the virial mass of the
halo, $r_s=R_v/c$ is the scale radius of the halo,
$C=\ln(1+c)-c/(1+c)$ and $c(M,Z)$ is the concentration parameter
\citep{munoz11}.   

\begin{figure}
\begin{center}
\includegraphics[width=8.5cm,angle=0.0 ]{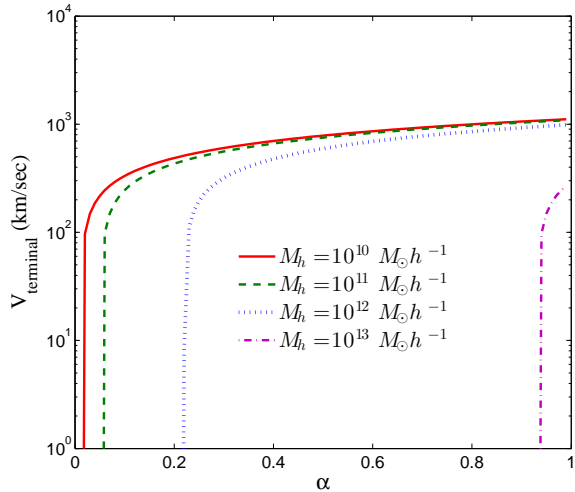}
 \caption {Terminal wind velocity as a function of energy injection
   efficiency $\alpha$, due to stellar feedback processes,
   for galactic haloes in the mass range, $10^{10-13} M_{\odot}
   h^{-1}$ (shown in different colors), at redshift, $z=0$.} 
  \label{fig-vterm}
 \end{center}
\end{figure}

\begin{figure*}
\begin{center}
\includegraphics[width=8.5cm,angle=0.0 ]{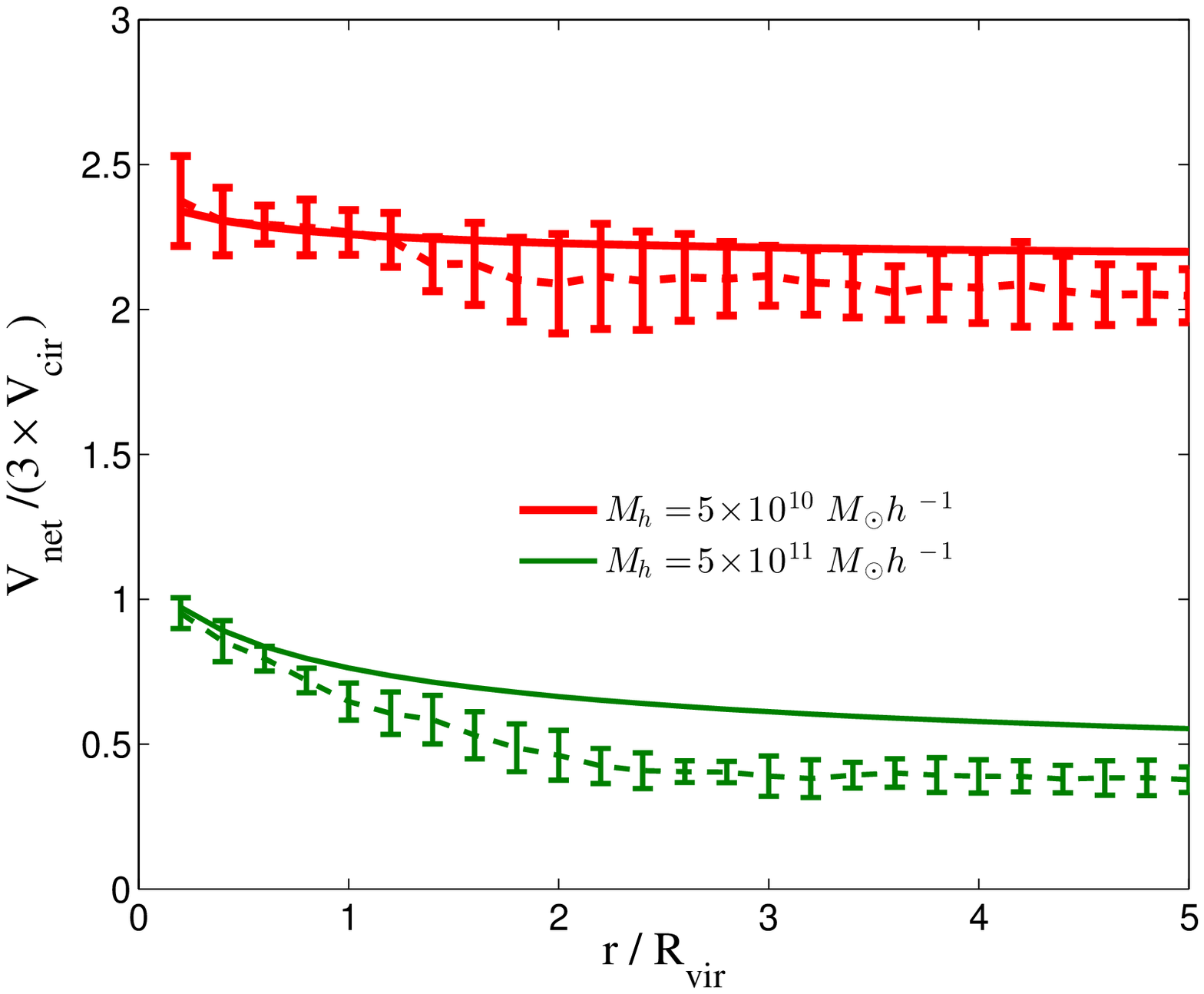}
\includegraphics[width=8.5cm,angle=0.0 ]{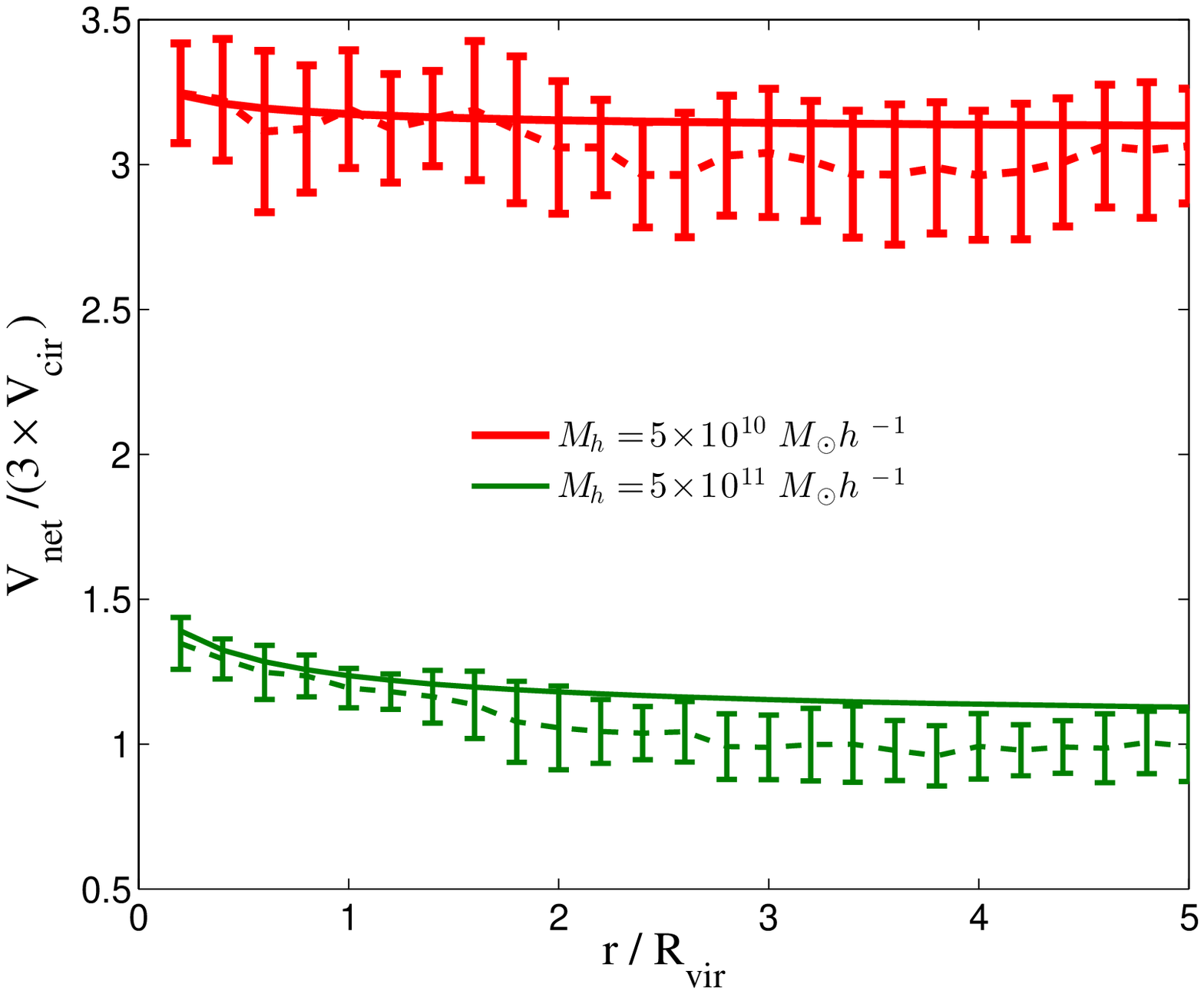}
\caption {Ratio of outflow velocity to $3\times$circular
   speed of the halo as a function of $r/R_{\rm{vir}}$ at z=3.0 (left panel) and z=1.0 (right panel).
   The thick red (thin green) solid line represents the unsuppressed wind velocity 
   whereas the dashed line represents
   the net outflow velocity (wind velocity-infall velocity) for halo mass $M_h=5\times 10^{10} M_{\odot} h^{-1}$
   ($M_h=5\times 10^{11} M_{\odot} h^{-1}$) at corresponding redshifts.} 
  \label{fig-avg}
 \end{center}
\end{figure*}

In Figure-\ref{fig-vterm}, we show the terminal wind velocity as a
function of $\alpha$, for the halo mass range $\sim
10^{10}$-$10^{13}h^{-1}M_{\odot}$, at z=0. 
For a given halo mass and redshift, the terminal wind velocity is
close to zero below a particular value of $\alpha$ ($v_{*}<v_{s}$),
beyond which there is a sharp increase in the wind velocity
($v_{*}>v_{s}$) and with further increase in $\alpha$, the terminal
velocity varies slowly ($v_{\rm{term}} \propto \sqrt{\alpha}$). 

Recent hydrodynamical simulations have shown that the efficiency of
energy deposition by multiple SNe can be as large as $\approx 0.3$,
which signifies that the rest of the energy is lost in radiation
\citep{sharma2014,vasiliev2015}.
These studies have investigated the radiative energy loss in the case
of SNe that are separated in time and in space, but are coherent
enough to mildly compensate for the radiative loss.
This is also corroborated by inference from X-ray observations of
outflows from M82 by \cite{strickland2009}. 
We fix the value of $\alpha$ at $0.3$ for the rest of this work.

\subsection{Infall velocity}
\label{subs-vin}

We use gravity only simulations run with the TreePM code
\citep{bagla02, khandai09} to compute the velocity of infalling gas
under the assumption that the gas particles follow the dark matter
particles.  
The suite of simulations used here is described in the
Table~\ref{table_nbody_runs}.  
The cosmological model and the power spectrum of fluctuations corresponds to
the best fit model for WMAP-5:
$\Omega_{nr} = 0.26$, $\Omega_{\Lambda} =0.74 $, $n_s = 0.96$, 
$\sigma_8 = 0.79$, $h=0.72$, $\Omega_b h^2 = 0.02273$
\citep{2009ApJS..180..330K}. 

We use the Friends-of-Friends (FOF) \citep{1985ApJ...292..371D}
algorithm with a linking length $l=0.2$ to identify haloes and construct a
halo catalog. 
Velocity field around each halo is obtained from the same simulations. 

The velocity field in the vicinity of haloes is highly anisotropic with
infall often along filaments and sheets.
We simplify the discussion here by considering only the radial motion
around haloes, and also by averaging in all directions around haloes.
This is an idealisation but should suffice to give us a glimpse of the
relative role of infall and outflows.

In order to calculate the infall speed (and hence the net outflow
speed), we divided the region around the halo into shells of thickness
$R_{\rm{vir}}/5$, where $R_{\rm{vir}}$ is the virial radius of the halo.
We calculate the average infall velocity of the shell by
averaging over the radial velocity of the particles present in the
shell. 
This gives the infall velocity of each shell as a function of distance
from the centre of the halo.  
We then average this radial infall velocity for approximately ten randomly selected
haloes for every mass scale and snapshot.
This gives the average radial infall velocity as a function of halo
mass and redshift.

%%%%%%%%%%%%%%%%%%%%%%%%%%%%%%%%%%%%%%%%%%%%%%%%%%%%%%%%%%%%%%%%%%%%%%
\begin{table}
  \caption{The table lists the simulations used here.  The first
    column lists the comoving size of the simulation box, the second
    column lists the minimum halo mass that we can resolve in the
    simulations, this is given in units of solar mass.  Each
    simulation was run with $512^3$ particles.  Cosmological
    parameters used here are described in text.} 
\label{table_nbody_runs}
\begin{center}
\begin{tabular}{||c|c||}
\hline
\hline
$L_{\rm{box}}$ (Mpc)& $M_{\rm{min}}$ ($M_{\odot}$)\\ 
\hline
\hline
$51.2$~h$^{-1}$ & $10^{9.02}$\\
\hline
$76.8$~h$^{-1}$ & $10^{9.43}$\\
\hline
$153.6$~h$^{-1}$ & $10^{10.33}$\\
\hline
\hline
\end{tabular}
\end{center}
\end{table}
%%%%%%%%%%%%%%%%%%%%%%%%%%%%%%%%%%%%%%%%%%%%%%%%%%%%%%%%%%%%%%%%%%%%%%

\section{Suppression of outflows}
\label{sec-supp}
\subsection{Suppression by infall}
\label{subs-infall}

Subtracting the infall velocity from the radial outflow velocity
gives the net outflow velocity.
This approach ignores the effect of pressure and assumes a pure advection
of outflow in the velocity field.
Thus our estimate of the effect of infall on outflows is likely to be
an under-estimate. 

In Figure- \ref{fig-avg}, we show the variation of the wind velocity and the
net outflow velocity (wind velocity-infall velocity)
as a function of the distance from the
centre of the dark matter halo.
We plot the ratio of distance from the centre to the virial radius of
the halo on the x-axis, and  the ratio of net outflow speed to
$3\times$circular speed of the halo along the y-axis.  
We also show the root-mean-square error on the net outflow velocity.
The main features of these plots are as follows:
\begin{itemize}
\item
  For a given redshift, the effect of infall increases with the
  increasing halo mass.
  This is mainly due to the increase in the gravitational field of the
  galaxy with its increasing mass, resulting in higher infall velocity. 
\item
  For a given halo mass, the effect of infall increases with the
  increasing redshift.
  This behaviour is due to the hierarchical formation history of the
  universe.
  Small galaxies   form early in the universe at high $\sigma$-peaks. 
  These galaxies then grow through accretion and merger to form larger
  galaxies and   galaxy clusters.
  The same halo mass corresponds to higher $\sigma$-peaks resulting in higher 
  infall velocity at higher redshifts.  
  \end{itemize}
Thus, by neglecting the effect of infall one may over-predict the
outflow velocity and hence the mass outflow rate.

\begin{figure}
\begin{center}
\includegraphics[width=9.0cm,angle=0.0]{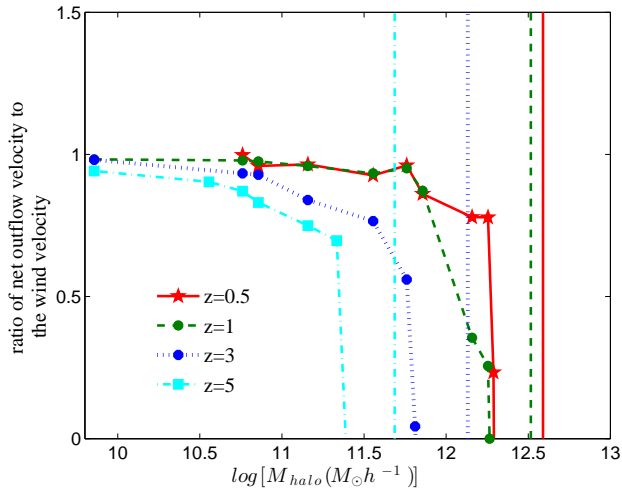}
 \caption {Ratio of net outflow velocity with and without the infall,
   as a function of the halo mass, for different redshifts (shown in
   different colors). The vertical lines represent the mass limit
   above which the outflows cannot overcome the gravitational field of
   the galaxy and the galaxies do not host the outflows even in the
   absence of the infall.}    
 \label{fig-netratio}
\end{center}
\end{figure}

In Figure-\ref{fig-netratio} we show the ratio of the net outflow velocity
with and without taking into account the effect of infall.
To find this ratio, we calculate the average radial velocity of the
shells between $R_{\rm{vir}}$ and 2$R_{\rm{vir}}$, for different redshifts.
For a given mass, the suppression of outflow due to the presence of
infall increases with increasing redshift. 
This effect is more prominent for high mass haloes as compared to low
mass haloes.
For example, haloes with $M_h \sim 10^{10} h^{-1} M_{\odot}$, the
difference in the suppression of the outflow due to infalling gas is
$<$10\% in the redshift range 0.5-5.0, whereas this difference
increases to 20\% at $M_h \sim 10^{11} h^{-1} M_{\odot}$ and 50\% at
$M_h \sim 2\times10^{11} h^{-1} M_{\odot}$.
This variation is due to the decrease in the outflow speed and
increase in infall speed with increasing halo mass.
This results in the sharp decline in the net outflow speed near
$M_{\rm{max}}$, where $M_{\rm{max}}$ is the upper mass limit beyond which there
are no effective outflows, as predicted by eqn \ref{terminal}.
The vertical lines in Figure-(\ref{fig-netratio}) show the values of
$M_{\rm{max}}$ at different redshift. 
The curves in the figure shows that the infall effectively suppresses
the outflows even for mass lower than $M_{\rm{max}}$, effectively
decreasing the value of $M_{\rm{max}}$, beyond which outflows cannot reach
the IGM. 
We find that the value of $M_{\rm{max}}$ decreases nearly by a factor of $2$
due to the presence of infall.  

Next, in Figure-\ref{fig-upmass} we show $M^{\rm{infall}}_{\rm{max}}$, the upper mass
limit beyond which there are no effective outflows as function of
redshift.  
$M^{\rm{infall}}_{\rm{max}}$ includes the effect of suppression of outflows due the
presence of infall which decreases the value of $M_{\rm{max}}$ approximately by a factor of two, independent of the 
redshift, as shown in
Figure-\ref{fig-netratio} and to illustrate this effect, we compare $M_{\rm{max}}$ (thin dashed, brown line)
with $M^{\rm{infall}}_{\rm{max}}$ as a function of redshift.
In this figure, we also show the value of $M^{\rm{infall}}_{\rm{max}}$ when the
infall velocity is calculated analytically from a spherical top hat
model (thin solid, cyan line).
It is interesting to find that the prediction of $M^{\rm{infall}}_{\rm{max}}$ from
N-body simulation (dot-dashed, blue line) and spherical collapse model agree well
with each other.  

\subsection{Suppression by hot CGM}
\label{subs-cgm}

\begin{figure}
\begin{center}
\includegraphics[width=8.5cm,angle=0.0]{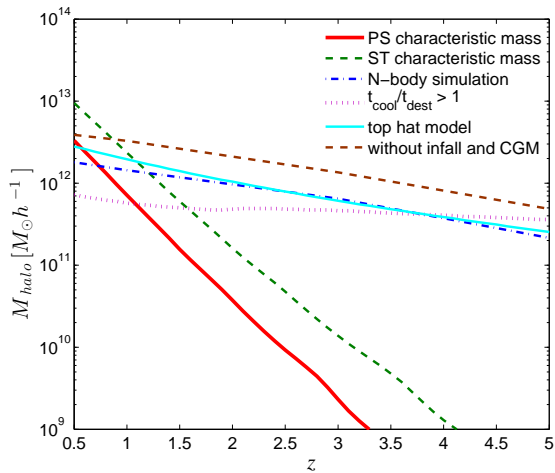}
 \caption {Comparison between $M_{\rm{max}}$ (thin dashed, brown line) which does not include the suppression of outflows
 by infall and CGM, 
 $M^{\rm{infall}}_{\rm{max}}$ calculated for N-body
   simulation (dot-dashed, blue line) and the top hat model (thin
   solid, cyan line), $M^{\rm{CGM}}_{\rm{max}}$ (dotted, pink line) and
   $M_{\rm{char}}$ calculated for the PS (thick solid, red line) and ST
   (dashed, green line) mass function.} 
 \label{fig-upmass}
\end{center}
\end{figure}

In Figure-\ref{fig-upmass}, we also show $M^{\rm{CGM}}_{\rm{max}}$ (dotted, pink
line), which is the mass limit above which the hot CGM exists in the
galactic halo.
This mass limit is determined by the condition
$\frac{t_{cool}}{t_{dest}}>1$, where $t_{cool}$ is the halo gas
cooling time and $t_{dest}$ is the timescale in which a halo forms a
larger halo through merger or accretion and the halo gas is reheated
during the process \citep{singh15a}. 
The gas cooling timescale is given by $t_{\rm cool}=3n_p k T/(2n^2_e \Lambda(Z,T))$, where $T$ is the gas temperature
(assumed to be the virial temperature of the galaxy), $n_e$ is the electron density (computed assuming that the CGM
contains approximately 10\% of the total halo mass), $n_p(\sim \mu_e n_e/\mu)$ is the total particle density with $\mu$
and $\mu_e$ the mean molecular weight of the gas and per free electron respectively and $\Lambda(Z,T)$ is the cooling
function \citep{dopita93} which depends on gas temperature and metallicity. The metallicity of the 
CGM is assumed to be $\sim 0.1 Z_{\odot}$.
The value of $M^{\rm{CGM}}_{\rm{max}}$ is consistent with earlier studies by
\cite{birn03} and \cite{keres05}. 
It changes only slightly with redshift and is comparable to
$M^{\rm{infall}}_{\rm{max}}$. 

Recent simulations have found that for the haloes more massive than
$M^{\rm{CGM}}_{\rm{max}}$, the outflow speed is reduced to the sound speed in the
CGM \citep{sarkar2015} due to the presence of hot environment around
the central galaxy.
This decelerates the outflow and even turns it around to form 
a galactic fountain.

The observations by \cite{mathes14} showed that the fraction of clouds
escaping the galactic halo decreases with increasing halo mass. 
They studied a sample of 14 galaxies in the redshift range $0.1<z<0.7$
using the quasar absorption spectroscopy. 
The authors divided the galaxy sample into two mass bins: massive
galaxies with $M_h>10^{11.5} M_{\odot}$ and the low mass galaxies with
$M_h<10^{11.5} M_{\odot}$.
The cloud escape fraction (within the virial radius of the galaxy) for
the low mass galaxies is $\sim 55$\% whereas the escape fraction
decreases significantly to $\sim 5$\% for massive galaxies.
It is interesting to find that the dividing mass limit between the two
samples is comparable to $M^{\rm{CGM}}_{\rm{max}}$ determined in the present
work.  
This observation supports the scenario of differential wind recycling
where the hot halo gas in massive galaxies decelerates the outflowing
clouds. For low mass  galaxies $M_h \lesssim 10^{12} M_{\odot}$, the
CGM cooling time is short (compared to halo destruction time), and,
consequently, the gas pressure is not sufficient to support against
gravity. As a result, the gas forms clumps, reducing its covering
fraction and hence the efficiency to suppress outflows. 

We note that, in addition to interfering with outflows, hot CGM forms
a barrier in front of the infalling gas.  
In a recent study by \cite{gabor15}, the authors found the quenching
of star formation due to the suppression of direct supply of the gas
to the central galaxy, in the mass range $10^{12}-10^{13} M_{\odot}$. 
The zoom-in hydrodynamic simulations by \cite{goerdt15} predict that
the gas infall velocity increases with increasing halo mass up to $M_h
\approx 10^{12} M_{\odot}$ beyond which the infall velocity decreases
as the halo mass is increased.
Note that, the halo mass where the relation between the infall
velocity and the halo mass is reversed, is independent of the galaxy
redshift and its comparable to $M^{\rm{CGM}}_{\rm{max}}$.
This indicates that the presence of hot CGM decreases the infall
velocity.
However, \cite{nelson15} found a significant suppression of the infall
by the galactic winds at small r ($< 0.5 R_{\rm{vir}}$) but they did not
find any change in accretion properties as a function of halo mass, in
the mass range $10^{10}$-$10^{12} M_{\odot}$.
Note that this is still not in disagreement with differential wind
recycling scenario if one uses $M_h \sim 10^{12} M_{\odot}$ as a
dividing line between the galaxies with and without a hot
circumgalactic environment.

\begin{figure}
\begin{center}
 \includegraphics[width=8.5cm,angle=0.0]{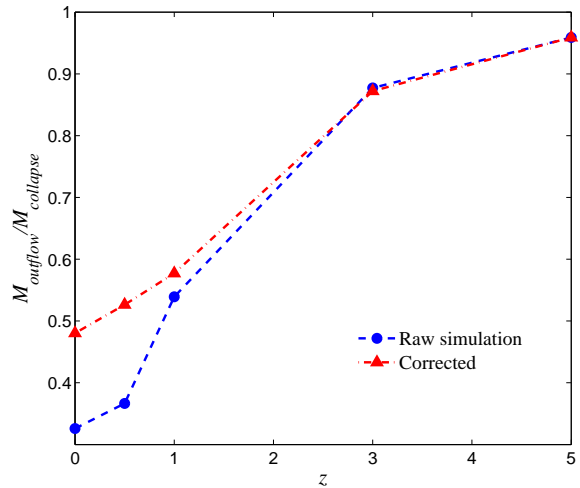}
 \caption {The ratio of mass in haloes hosting unsuppressed outflows
   to the total mass in all collapsed structures.  The blue curve
   (dashed) shows the ratio as determined in N-Body simulations.  Given that we
 use simulations with different mass resolution and we are unable to
 resolve low mass haloes in simulations used at low redshifts, this
 approach under-estimates the ratio.  The red curve (dot-dashed) is obtained by
 applying a correction by extrapolating the mass function to the same
 mass resolution as is available in the simulations used for high
 redshift snapshots.  The qualitative trend remains the same though
 the ratio has the value closer to $0.5$ rather than $0.3$.} 
 \label{fig-mout}
\end{center}
\end{figure}

\subsection{Relative importance of infall versus hot CGM in
  suppressing outflows} 
\label{subs-inf-cgm}

The cosmological infall and the presence of hot CGM, both give the
upper mass limits,  $M^{\rm{infall}}_{\rm{max}}$ and $M^{\rm{CGM}}_{\rm{max}}$
respectively, beyond which the outflow is halted.
If $M^{\rm{infall}}_{\rm{max}} < M^{\rm{CGM}}_{\rm{max}}$, the outflow is suppressed by
the infall, even before the existence of the pressure supported, hot
circumgalactic gas.
In this scenario, the infall plays more important role than the CGM in
suppressing the outflows.
However, if $M^{\rm{infall}}_{\rm{max}} > M^{\rm{CGM}}_{\rm{max}}$, the galaxies host the
hot CGM before the infall velocity becomes comparable to the outflow
velocity.
Therefore, the process corresponding to minimum of the two
mass-limits, dominants the suppression.  
Comparing $M^{\rm{infall}}_{\rm{max}}$ and $M^{\rm{CGM}}_{\rm{max}}$ (see figure-\ref{fig-upmass}) suggest that
at low redshifts ($z\lesssim3.5$), the role of hot CGM, in suppressing the
outflows is more important than the infall, since $M^{\rm{infall}}_{\rm{max}} >
M^{\rm{CGM}}_{\rm{max}}$, whereas at high redshifts where $M^{\rm{infall}}_{\rm{max}} <
M^{\rm{CGM}}_{\rm{max}}$($z>3.5$), infall suppresses the outflow more
effectively than the CGM.
The two mass limits are close in the redshift range considered.
$M^{\rm{infall}}_{\rm{max}}$ is larger than $M^{\rm{CGM}}_{\rm{max}}$
by a factor of $2-3$ near $z \sim 0$. The difference between the two mass-limits
decreases with increasing redshift. 
Therefore, these two processes give a mass range separating the haloes
with and without effective outflows. 

\section{Fraction of galaxies affected by the suppression of outflows}
\label{sec-igm}

The intergalactic medium is enriched by  metals ejected by galaxies
through galactic winds.
Therefore, the enrichment level of IGM crucially depends on the
feedback mechanism, which throws out metals into the IGM and the
processes such as infall and the existence of hot CGM, which suppress
the outflows.
In this section, we examine whether the suppression of outflows by
cosmological infall/hot CGM affects a significant fraction of the
galaxy population. 

In Figure-\ref{fig-upmass}, we also show the characteristic mass
$M_{\rm{char}}$ for the Press-Schechter (PS) (defined as $\frac{\delta^2(z)}{2 \sigma^2(M_{\rm{char}})}=1$)
as well as Sheth-Tormen (ST) mass function ($\alpha_{\rm{ST}} \frac{\delta^2(z)}{2 \sigma^2(M_{\rm{char}})}=1$,
 where $\alpha_{\rm{ST}}=0.707$).  
The characteristic mass, $M_{\rm{char}}$ represents the mass scale beyond
which the number of haloes decreases rapidly with increasing halo
mass.
Therefore, the haloes with masses below $M_{\rm{char}}$, dominate the halo
population.
The present day value of $M_{\rm{char}} \sim 1.4\times10^{13} h^{-1}
M_{\odot}$ for Press-Schechter mass function and it decreases with
increasing redshift as expected from the hierarchical structure
formation scenario.
We find that at $z > 1-2$, $M^{\rm{infall}}_{\rm{max}}$ and $M^{\rm{CGM}}_{\rm{max}}$
$>M_{\rm{char}}$, which implies that the haloes in which outflows are
completely suppressed by infalling gas or the presence of the hot CGM,
are rare. 
At low redshifts, majority of the haloes lie near the characteristic
mass range, and  therefore outflow suppression becomes important. 
Therefore, the suppression of the outflows due the infall as well as
the hot CGM should be taken into account while dealing with the haloes
hosting outflows, especially at low redshifts.

Next, to get an estimate of the population of galaxies with suppressed
or unsuppressed outflows, we compute the ratio of the total mass in the
galaxies hosting unsuppressed outflows to the total mass present in
collapsed structures.
In Figure-\ref{fig-mout}, we show this ratio as a function of redshift.
The total mass in the collapsed structures is estimated using the
N-body simulation.  
To compute the mass in the haloes supporting outflows, we consider
haloes with $M_h \lesssim Min[M^{\rm{infall}}_{\rm{max}}, M^{\rm{CGM}}_{\rm{max}}$].
Note that, while calculating the number of galaxies with unsuppressed
outflows, we exclude those haloes in which outflows are
{\it{completely}} suppressed.
Therefore, the ratio includes galaxies with partially suppressed
outflows. 

Given that the mass resolution of simulations used in this study is
different, and that simulations with a larger box-size are required
for low redshift studies \citep{2006MNRAS.370..993B}, the mass
resolution is not independent of redshift in the analysis.
The mass resolution is poorer at low redshifts and therefore we miss
out on low mass haloes that can potentially contribute to outflows.
To overcome this problem, we compute the mass function at each
redshift and extrapolate it to the mass resolution of the highest
resolution simulation used here.
The red curve (dot-dashed) in figure-\ref{fig-mout} incorporates this correction while computing
the ratio of total mass in galaxies that have an unsuppressed outflow
to the total mass in collapsed haloes.

The ratio is close to unity at high redshifts ($z\gtrsim 5$) due to
the hierarchical formation history of the universe, the haloes
at high redshifts are mostly low mass ones.
The fraction of mass present in the outflow-supporting haloes
decreases with decreasing redshift due to the formation of massive
haloes at low redshifts.  
The redshift range, $z \sim 1-2$, represents the era of high star
formation activity.
Hence, the abundance of outflow supporting systems can potentially
determine the enrichment history of the galaxies, CGM and IGM.
We find that the ratio $M_{outflow}/M_{collapse}$ decreases from 70\%
to 60\% in this redshift range.
Thus, a small but significant fraction of haloes lie in the mass 
range where the outflows are completely suppressed by the infall or the presence
of the hot CGM during the era of high star formation hence high
feedback activity.
In the redshift range $z < 1$, $M_{outflow}/M_{collapse}$ decreases
from 60\% to less than 50\%. 
Since, the presence of the hot CGM is more important at low redshifts
(see figure-\ref{fig-upmass}), the outflowing gas in approximately half of
the haloes at low redshift, is decelerated and suppressed by the
surrounding CGM in these galactic haloes.

Therefore, the enrichment of the IGM is easier at high redshifts
whereas at low redshifts, the metal carrying outflows are suppressed
more effectively, mostly by the hot CGM.
These suppressed outflows are then recycled into the  galaxy hence
enriching the CGM as well as the stars.
This result is in agreement with the simulation result of
\cite{dave07}, indicating that the IGM contains more metals at high
redshifts whereas the stars and the halo gas contain more metals at
low redshifts. 

\section{Caveats}
\label{sec-limit}
The usefulness of our results discussed above is, however, subject to certain caveats.
Firstly, the geometry of the infalling gas near the virial radius may not be isotropic as we have assumed, and
is likely to be in the form of streams \citep{dekel09a, dekel09b}. Simulations show that gas mainly flows along
filaments in the cosmic web, and the enhanced density of gas in the filaments causes it to cool, and
avoids being shocked to the virial temperature \citep{birn03}.
Considerations of such cold streams with regard to the
fraction of outflowing gas that can escape is difficult without hydrodynamical simulations, and beyond the
scope of the present work. However, we note that the cold mode of accretion dominates galaxies below
the mass scale of $\sim 10^{12}$ M$_\odot$, and the mode of accretion changes to that of slow cooling from hot
halo gas (hot mode) beyond this mass scale \citep{keres05}. The shock heating and disruption of cold infalling streams, 
especially at high redshifts and low galacto-centric radii, becomes more important near M$_h\sim 10^{12}$ M$_\odot$ 
\citep{danovich12, nelson13, danovich14, gabor14, nelson15b}. This mass scale coincides with the mass scale 
shown in Figure-\ref{fig-upmass} (dotted pink line corresponding to the galactic haloes 
with $t_{\rm cool}/t_{\rm dest}>1$),
and therefore the filamentary nature of cold flows in low mass galaxies does not significantly alter our conclusions.

Secondly, all of the CGM gas may not be in a hot, diffuse state, and a fraction of it is likely
to be in a warm ($\sim 10^4$ K) phase, as indicated by the COS-Halos survey \citep{werk2014}.
It is also believed that the interaction of the CGM gas with the outflows driven by first phases of star formation
may cause clumping \citep{marinacci2010, sharma2014}. The gas in the interaction zone may suffer from
various instabilities, such as thermal instability (due to the mixing of gas at different temperatures
and densities) and Kelvin-Helmholz instability (due to shear). The resulting structures and turbulence
 in the CGM may allow
some fraction of the outflowing gas to escape, in the cases where our results above for a homogeneous 
CGM may not allow any escape. However, this fraction is difficult to estimate without the aid of hydrodynamic
simulations.

However, given these uncertainties, it is interesting to note the similarities of the mass scale we have
discussed so far, that of $M_{\rm max}^{\rm infall}$, with other mass scales that are significant for galaxy
evolution. Several studies (e.g, \cite{Behroozi2013}) have shown that the ratio of baryonic mass to the total
mass of galaxies reaches a maximum around $\sim 10^{12}$ M$_\odot$ at the present epoch, and slightly
lower at high redshift (e.g., lower by a factor of $\sim 3$ at $z\sim 4$). This is remarkably close to the mass
scale $M_{\rm max}^{\rm infall}$ shown in Figure-\ref{fig-upmass} (dot-dashed blue line and thin solid cyan line). 
We can speculate, on the basis of our
calculations here, that the stoppage of the outflow is causally connected to the baryon-to-total mass ratio. 
The analytic work of \cite{sharma2014}, which did not consider the effect of infall or the presence of CGM, and
only considered the effect of gravity, has already suggested that the stoppage of outflow is related to the
baryons-to-total mass ratio. Our work on the effect of infall and CGM's presence provides additional supports for this
scenario.

\section{Conclusions}
\label{sec-conc}
In this work, we have studied the relative importance of galactic
outflows, cosmological infall and the role of presence of hot
circumgalactic gas on these processes under some simplified
assumptions such as spherically symmetric infall, constant value of
energy injection efficiency, and gas particles following the dark
matter particles. We have also neglected the complexities involved in the dynamics and distribution
of the CGM as well as the infalling gas. However, as discussed in section-\ref{sec-limit}, the 
detailed treatment of these complexities is unlikely to change the conclusions drawn from
our analysis substantially.
Our conclusions can be summarised as follows: 

\begin{enumerate}
\item 
  Without considering the CGM, the infalling gas interferes with the
  outflowing gas reducing the net outflow velocity. 
  This reduction depends on the mass and redshift of the galaxy.
  The larger (massive) galaxies at high redshifts suffer more
  suppression compared to smaller (low mass) galaxies at low redshifts
  due to the combined effect of weaker outflows and stronger infall in
  the case of massive galaxies. 
\item
  Even in the absence of any infall, there exits an upper mass limit
  beyond which the outflows are unable to overcome the gravitational
  field of  the galaxy.
  This upper mass limit decreases roughly by a factor of two due to
  the additional suppression of outflows by infalling gas, independent
  of the redshift.
  The value of this upper mass limit, $M^{\rm{infall}}_{\rm{max}}$, predicted by
  simulations agrees well with the value predicted by the top-hat
  model. 
\item
  In addition to the infall, the presence of the hot gaseous
  environment in the form of CGM, decelerates the outflows.
  The CGM also gives an upper mass limit, $M^{\rm{CGM}}_{\rm{max}}$, beyond
  which the hot CGM effectively stops the outflows from escaping the
  galaxy.  
  The hot CGM may also reduce the direct supply of infalling gas to
  the central part of the galaxy. 
  We estimate $M^{\rm{CGM}}_{\rm{max}}$ under the condition that the gas cooling
  time exceeds the halo destruction timescale.
  $M^{\rm{CGM}}_{\rm{max}}$ varies slowly with the redshift with its value $\sim
  10^{12} M_{\odot}$ in the redshift range $0-5$.
\item 
  The hot CGM is more effective than the infall in the low redshift
  range 0-3.5 in counteracting the outflows whereas the infall becomes
  more effective at high redshift ($z>3.5$).
  The upper mass limits, for suppressing the outflows, predicted by
  both the processes are comparable and together determine the fate of
  the outflowing gas. 
\item 
  Comparison of $M^{\rm{CGM}}_{\rm{max}}$ and $M^{\rm{infall}}_{\rm{max}}$ with the
  characteristic mass predicts the suppression of the outflows to be
  important at low redshifts ($z<1-2$), where the galaxies with
  completely suppressed outflows constitute a significant fraction of
  the overall galaxy population. 
  The fraction of galaxies with unsuppressed outflows predicted by the
  simulation decreases from $\sim$90\% at $z=5$ to $\sim$50\% at
  $z=0$. This fraction is of order  60-70\% in the era of high star
  formation and hence high feedback activity ($z\sim 1-2$), likely
  affecting the enrichment history of the universe.  
\end{enumerate}
\vspace{10mm}
{\bf{ACKNOWLEDGEMENTS}}\\
We thank the anonymous referee for valuable comments and suggestions which helped in improving this paper.
We also thank Girish Kulkarni and Kartick C. Sarkar for helpfull discussions. PS acknowledges
the hospitality of IISER, Mohali.  

\footnotesize{
\bibliography{bibtexinfall}{}
\bibliographystyle{mn2e}
}

\end{document}